# Near-field thermal electromagnetic transport: An overview

Sheila Edalatpour[†], John DeSutter and Mathieu Francoeur[†]

*Radiative Energy Transfer Lab, Department of Mechanical Engineering, University of Utah, Salt Lake City, UT 84112, USA*

**ABSTRACT**

A general near-field thermal electromagnetic transport formalism that is independent of the size, shape and number of heat sources is derived. The formalism is based on fluctuational electrodynamics, where fluctuating currents due to thermal agitation are added to Maxwell's curl equations, and is thus valid for heat sources in local thermodynamic equilibrium. Using a volume integral formulation, it is shown that the proposed formalism is a generalization of the classical electromagnetic scattering framework in which thermal emission is implicitly assumed to be negligible. The near-field thermal electromagnetic transport formalism is afterwards applied to a problem involving three spheres with size comparable to the wavelength, where all multipolar interactions are taken into account. Using the thermal discrete dipole approximation, it is shown that depending on the dielectric function, the presence of a third sphere slightly affects the spatial distribution of power absorbed compared to the two-sphere case. A transient analysis shows that despite a non-uniform spatial distribution of power absorbed, the sphere temperature remains spatially uniform at any instant due to the fact that the thermal resistance by conduction is much smaller than the resistance by radiation. The formalism proposed in this paper is general, and

---

[†] Corresponding authors. Tel.: +1 801 581 5721, Fax: +1 801 585 9825

E-mail addresses: mfrancoeur@mech.utah.edu (M. Francoeur), sheila.edalatpour@utah.edu (S. Edalatpour)

could be used as a starting point for adapting solution methods employed in traditional electromagnetic scattering problems to near-field thermal electromagnetic transport.

## 1. INTRODUCTION

Thermal radiative transport problems are typically analyzed using Planck's theory of heat radiation in which two physical mechanisms are omitted [1]. First, radiation transport is treated as incoherent (rays or photons) such that wave interference is neglected. Additionally, evanescent modes, decaying exponentially within a distance of approximately a wavelength normal to the surface of a thermal source, are not taken into account. Neglecting wave interference and heat transfer by tunneling of evanescent modes is reasonable as long as the size of the bodies and their separation distance is much larger than the thermal wavelength, which is approximately 10 μm at room temperature. When the size of the bodies and/or their separation distance is comparable to or smaller than the wavelength, Planck's theory ceases to be valid and radiation heat transfer is said to be in the near-field regime. The most remarkable near-field effect is the enhancement of radiative heat exchange beyond Planck's blackbody limit due to tunneling of evanescent modes. Near-field thermal radiation problems are typically modeled via fluctuational electrodynamics in which the Maxwell curl equations are augmented by fluctuating current sources due to thermal agitation [2]. These fluctuating currents are distributed throughout the volume of a heat source and are related to its local temperature via the fluctuation-dissipation theorem that is applicable under the assumption of local thermodynamic equilibrium. Wave interference and tunneling of evanescent modes are, therefore, taken into account when applying fluctuational electrodynamics. Despite the fact that fluctuational electrodynamics is a phenomenological theory, the validity of this formalism has been demonstrated experimentally for separation distances from a few micrometers down to a few tens of nanometers [3-10].

The growing interest in near-field thermal electromagnetic transport is driven by numerous potential applications to thermophotovoltaic power generation [11-18], localized radiative cooling [19], nanomanufacturing [20], thermal emission control [21-23] and thermal rectification [24-27]. In terms of modeling, closed-form solutions of near-field thermal electromagnetic transport have been derived for special geometries such as one-dimensional layered media [28,29], two large spheres [30-32], a sphere and a surface [33,34] and an arbitrary number of nanoparticles modeled as electric point dipoles [35,36] using the method of dyadic Green's function. Moreover, a number of numerical methods have been adapted to near-field thermal radiation: the finite-difference time-domain method [37-39], the finite-difference frequency-domain method [40], the boundary element method [41], the method of moments [42] and the discrete dipole approximation, which has been referred to as the thermal discrete dipole approximation (T-DDA) [43,44]. Additional information about near-field thermal radiation modeling and its potential engineering applications can be found in the numerous review papers published during the past decade [28,45-52].

The objective of this overview paper is to provide a general near-field thermal electromagnetic transport formalism that is applicable within the limit of validity of fluctuational electrodynamics [53]. The proposed formalism is essentially a generalization of the classical electromagnetic scattering theory [54] in which thermal emission is non-negligible. As such, a secondary objective of this work is to provide a bridge between the electromagnetic scattering and near-field thermal radiation theories that are otherwise treated as two distinct fields. This is particularly important since thermal effects are usually ignored in electromagnetic scattering problems, while their contribution might be considerable in applications such as nanoparticle patterning with a laser [20,55]. The rest of the paper is organized as follows. A volume integral equation for the total electric field is derived starting from the Maxwell curl equations augmented by fluctuating currents. It is shown that when thermal emission is negligible, the expression for the total electric field reduces to the volume integral equation used in traditional electromagnetic scattering. Afterwards, a near-field radiative heat transfer problem involving three spheres is solved starting from the volume integral equation for the total electric field and by applying the T-DDA, which is an approach adapted from the electromagnetic scattering literature [56,57]. Concluding remarks are provided in the last section.

## 2. NEAR-FIELD THERMAL ELECTROMAGNETIC TRANSPORT FORMALISM

Electromagnetic transport is modeled via the macroscopic Maxwell equations. Assuming exp(-$i\omega t$) for the time-harmonic fields, the Maxwell curl equations in the frequency domain are:

$$\nabla \times \mathbf{E}(\mathbf{r},\omega) = i\omega\mu\mu_0 \mathbf{H}(\mathbf{r},\omega) \tag{1}$$

$$\nabla \times \mathbf{H}(\mathbf{r},\omega) = -i\omega\varepsilon\varepsilon_0 \mathbf{E}(\mathbf{r},\omega) \tag{2}$$

where $\mathbf{E}$ and $\mathbf{H}$ are the time-independent electric and magnetic fields, $\mathbf{r}$ is the position vector where the fields are observed, $\omega$ is the angular frequency, $i$ is the complex constant, $\varepsilon_0$ and $\mu_0$ are the vacuum electric permittivity and magnetic permeability, while $\varepsilon$ and $\mu$ are the frequency-dependent dielectric function and relative magnetic permeability that are, in the most general case, complex numbers. The Maxwell equations as given above do not include thermal emission. Fluctuational electrodynamics, which is a phenomenological framework, is typically used to account for thermal emission [2,58]. In fluctuational electrodynamics, the thermal radiation field is conceptualized as an electromagnetic field generated by the stochastic oscillations of charges induced by thermal agitation. Mathematically, the stochastic oscillations of charges are modeled by fluctuating currents, due to electric ($\mathbf{J}^{fl,e}$) and magnetic ($\mathbf{J}^{fl,m}$) dipole oscillations, that are added to Maxwell's curl equations:

$$\nabla \times \mathbf{E}(\mathbf{r},\omega) = i\omega\mu\mu_0 \mathbf{H}(\mathbf{r},\omega) - \mathbf{J}^{fl,m}(\mathbf{r},\omega) \tag{3}$$

$$\nabla \times \mathbf{H}(\mathbf{r},\omega) = -i\omega\varepsilon\varepsilon_0 \mathbf{E}(\mathbf{r},\omega) + \mathbf{J}^{fl,e}(\mathbf{r},\omega) \tag{4}$$

The fluctuating currents $\mathbf{J}^{fl,m}$ and $\mathbf{J}^{fl,e}$ are fully described by their first two moments. The first moment of the fluctuating currents, which corresponds to their ensemble average, is zero. The second moment of the fluctuating currents, which is the ensemble average of their spatial correlation function, is given by the fluctuation-dissipation theorem [2]:

$$\left\langle \mathbf{J}^{fl,m}(\mathbf{r}',\omega) \otimes \mathbf{J}^{fl,m}(\mathbf{r}'',\omega') \right\rangle = \frac{4\omega\mu_0\mu''}{\pi}\Theta(\omega,T)\delta(\mathbf{r}'-\mathbf{r}'')\delta(\omega-\omega')\bar{\bar{\mathbf{I}}} \tag{5}$$

$$\left\langle \mathbf{J}^{fl,e}(\mathbf{r}',\omega) \otimes \mathbf{J}^{fl,e}(\mathbf{r}'',\omega') \right\rangle = \frac{4\omega\varepsilon_0\varepsilon''}{\pi}\Theta(\omega,T)\delta(\mathbf{r}'-\mathbf{r}'')\delta(\omega-\omega')\bar{\bar{\mathbf{I}}} \tag{6}$$

where $\otimes$ denotes the outer product, $\bar{\bar{\mathbf{I}}}$ is the unit dyadic, while $\varepsilon''$ and $\mu''$ are respectively the imaginary parts of the dielectric function and the relative magnetic permeability. The term $\Theta(\omega,T)$ is the mean energy of an electromagnetic state at frequency $\omega$ and temperature $T$ calculated as:

$$\Theta(\omega,T) = \frac{\hbar\omega}{\exp(\hbar\omega/k_B T) - 1} \tag{7}$$

where $\hbar$ and $k_B$ are respectively the Planck and Boltzmann constants. Note that the fluctuating currents due to electric and magnetic dipole oscillations are not spatially correlated [2]. The fluctuation-dissipation theorem, applicable to objects in local thermodynamic equilibrium, provides the necessary link between the electromagnetic representation of thermal emission and the local temperature $T$.

A general near-field thermal electromagnetic transport formalism is established starting from the thermal stochastic Maxwell equations (Eqs. (3) and (4)) and by using Fig. 1, where an arbitrary number of finite objects are embedded into an infinite medium. The infinite medium occupies a volume $V_1$ and is assumed to be homogeneous, linear, isotropic, non-magnetic, non-absorbing and thus non-emitting. The objects occupy as a whole a finite interior volume $V_2$. The entire three-dimensional space is given by $\Re^3 = V_1 \cup V_2$. It is assumed that the material of the interior region $V_2$ is isotropic, linear, possibly inhomogeneous, non-magnetic and in local thermodynamic equilibrium. While the vast majority of naturally occurring materials are non-magnetic at optical and infrared frequencies, man-made structures such as metamaterials may lead to effectively magnetic responses within the aforementioned spectral band [59-63]. Under the assumption that all media in $\Re^3$ are non-magnetic, the relative magnetic permeability $\mu$ in Eq. (3) is equal to unity while the fluctuating current due to magnetic dipole oscillations $\mathbf{J}^{fl,m}$ vanishes. Note that it would be straightforward to extend the formalism presented hereafter to magnetic materials in $V_2$. The remaining fluctuating current in Eq. (4), due electric dipole oscillations, will be referred to as $\mathbf{J}^{fl}$ in order to simplify the nomenclature. The electric field due to external sources, such as illumination by a laser or thermal emission from the surroundings (sometimes referred to as the thermal bath), is accounted for via an incident electric field $\mathbf{E}^{inc}$.

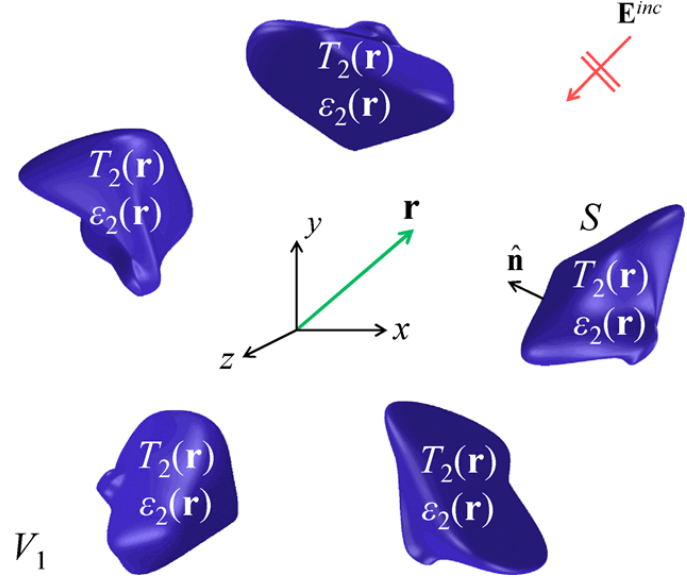

Figure 1. Graphical representation of a general near-field thermal electromagnetic transport problem. The interior region of total volume $V_2$ is composed of objects embedded into the exterior region of volume $V_1$. Individual objects may have different, inhomogeneous dielectric functions and temperatures. Illumination of the objects by external sources is represented by the incident electric field vector $\mathbf{E}^{inc}$.

The boundary conditions linking the electromagnetic fields across the interface $S$ delimiting the exterior and interior regions are given by:

$$\hat{\mathbf{n}} \times [\mathbf{E}_1(\mathbf{r},\omega) - \mathbf{E}_2(\mathbf{r},\omega)] = \mathbf{0}, \quad \mathbf{r} \in S \tag{8}$$

$$\hat{\mathbf{n}} \times [\mathbf{H}_1(\mathbf{r},\omega) - \mathbf{H}_2(\mathbf{r},\omega)] = \mathbf{0}, \quad \mathbf{r} \in S \tag{9}$$

where $\hat{\mathbf{n}}$ is a local outward unit vector normal to $S$, while the subscripts 1 and 2 refer to the exterior and interior regions, respectively. These boundary conditions imply that the tangential components of the electric and magnetic fields are continuous across the interface $S$.

From now on, the volume integral formulation of near-field thermal electromagnetic transport is adopted. This volume integral formulation is determined by first considering the vector wave equations for the electric field in the exterior and interior regions that are derived from the Maxwell curl equations:

$$\nabla \times \nabla \times \mathbf{E}(\mathbf{r},\omega) - k_1^2 \mathbf{E}(\mathbf{r},\omega) = \mathbf{0}, \quad \mathbf{r} \in V_1 \tag{10}$$

$$\nabla \times \nabla \times \mathbf{E}(\mathbf{r},\omega) - k_2^2 \mathbf{E}(\mathbf{r},\omega) = i\omega\mu_0 \mathbf{J}^{fl}(\mathbf{r},\omega), \quad \mathbf{r} \in V_2 \tag{11}$$

where $k_1$ is the magnitude of the wavevector in $V_1$ (real number), while $k_2$ is the magnitude of the wavevector in $V_2$, which is, in the general case, a complex number. Note that the magnetic field at any location in $\Re^3$ can be determined from the electric field using Eq. (3). Equations (10) and (11) can be combined into a single inhomogeneous equation that is applicable everywhere in $\Re^3$:

$$\nabla \times \nabla \times \mathbf{E}(\mathbf{r},\omega) - k_1^2 \mathbf{E}(\mathbf{r},\omega) = i\omega\mu_0 \mathbf{J}(\mathbf{r},\omega), \quad \mathbf{r} \in \Re^3 \tag{12}$$

where the current $\mathbf{J}$ is an equivalent source function. In the interior region, its expression is given by:

$$\mathbf{J}(\mathbf{r},\omega) = \mathbf{J}^{fl}(\mathbf{r},\omega) - \frac{i}{\omega\mu_0}(k_2^2 - k_1^2)\mathbf{E}(\mathbf{r},\omega), \quad \mathbf{r} \in V_2 \tag{13}$$

where the second term on the right-hand side of the equation is the source function for the scattered field. The equivalent current $\mathbf{J}$ vanishes in the exterior region due to the absence of scattering objects and since the material filling $V_1$ is non-emitting. The solution of the inhomogeneous linear differential equation (12) is split into two parts, namely a solution of the homogeneous equation and a particular solution of the inhomogeneous equation. The homogeneous equation is given by:

$$\nabla \times \nabla \times \mathbf{E}^{inc}(\mathbf{r},\omega) - k_1^2 \mathbf{E}^{inc}(\mathbf{r},\omega) = \mathbf{0}, \quad \mathbf{r} \in \Re^3 \tag{14}$$

where $\mathbf{E}^{inc}$ corresponds to the electric field that would exist in the absence of objects. The particular solution of Eq. (12) is the sum of the scattered and fluctuating fields generated by the equivalent current $\mathbf{J}$. The scattered and fluctuating fields, determined using the free-space dyadic Green's function (DGF) $\overline{\overline{\mathbf{G}}}$, must satisfy the boundary conditions given by Eqs. (8) and (9) as well as the radiation condition at infinity [54]. After some mathematical manipulations, the sum of the scattered and fluctuating electric fields is given by:

$$\mathbf{E}^{sca}(\mathbf{r},\omega) + \mathbf{E}^{fl}(\mathbf{r},\omega) = i\omega\mu_0 \int_{V_2} \overline{\overline{\mathbf{G}}}(\mathbf{r},\mathbf{r}',\omega) \cdot \mathbf{J}(\mathbf{r}',\omega) d^3\mathbf{r}', \quad \mathbf{r} \in \Re^3 \tag{15}$$

where $\overline{\overline{\mathbf{G}}}(\mathbf{r},\mathbf{r}',\omega) = \frac{e^{ik_1 R}}{4\pi R}\left[\left(1 - \frac{1}{(k_1 R)^2} + \frac{i}{k_1 R}\right)\overline{\overline{\mathbf{I}}} - \left(1 - \frac{3}{(k_1 R)^2} + \frac{3i}{k_1 R}\right)\hat{\mathbf{R}} \otimes \hat{\mathbf{R}}\right]$ (16)

In the above expressions, $\mathbf{r}'$ is a source point located in $V_2$, $R = |\mathbf{r} - \mathbf{r}'|$ and $\hat{\mathbf{R}} = (\mathbf{r} - \mathbf{r}')/|\mathbf{r} - \mathbf{r}'|$. The (total) electric field at $\mathbf{r}$ is determined by adding the incident field to Eq. (15):

$$\mathbf{E}(\mathbf{r},\omega) = i\omega\mu_0 \int_{V_2} \overline{\overline{\mathbf{G}}}(\mathbf{r},\mathbf{r}',\omega) \cdot \mathbf{J}(\mathbf{r}',\omega) d^3\mathbf{r}' + \mathbf{E}^{inc}(\mathbf{r},\omega), \quad \mathbf{r} \in \Re^3 \tag{17}$$

The volume integral equation (17) is a general expression describing near-field thermal electromagnetic transport. In the absence of objects, the (total) electric field is simply equal to the incident field. In the special case that thermal emission by the objects is negligible compared to the incident field, the fluctuating current $\mathbf{J}^{fl}$ vanishes and the volume integral equation for the electric field reduces to:

$$\mathbf{E}(\mathbf{r},\omega) = k_1^2 \int_{V_2} [(\varepsilon_2 / \varepsilon_1) - 1] \overline{\overline{\mathbf{G}}}(\mathbf{r},\mathbf{r}',\omega) \cdot \mathbf{E}(\mathbf{r}',\omega) d^3\mathbf{r}' + \mathbf{E}^{inc}(\mathbf{r},\omega), \quad \mathbf{r} \in \mathfrak{R}^3 \quad (18)$$

which is the volume integral equation used for solving classical electromagnetic scattering problems [54].

A near-field radiative heat transfer problem involving three spheres is solved next using the T-DDA and starting with the volume integral equation (17).

## 3. NEAR-FIELD RADIATIVE HEAT TRANSFER BETWEEN THREE SPHERES

As an illustrative example, a near-field radiative heat transfer problem involving three spheres is analyzed hereafter starting from the formalism described in the previous section. It is assumed that the exterior region is a vacuum ($\varepsilon_1 = 1$), such that the equivalent current is simplified as follows:

$$\mathbf{J}(\mathbf{r},\omega) = \begin{cases} \mathbf{0}, & \mathbf{r} \in V_1 \\ \mathbf{J}^{fl}(\mathbf{r},\omega) - i\omega\varepsilon_0(\varepsilon_2 - 1)\mathbf{E}(\mathbf{r},\omega), & \mathbf{r} \in V_2 \end{cases} \quad (19)$$

A schematic representation of the problem is shown in Fig. 2, where spheres *A*, *B* and *C*, of same diameter *D* and homogeneous dielectric function $\varepsilon_2$, are separated by distances *d*. It is assumed that only sphere *A* is emitting while both spheres *B* and *C* are pure absorbers ($T_B = T_C = 0$ K). The objective here is to analyze how the presence of sphere *C* affects the power absorbed within sphere *B*.

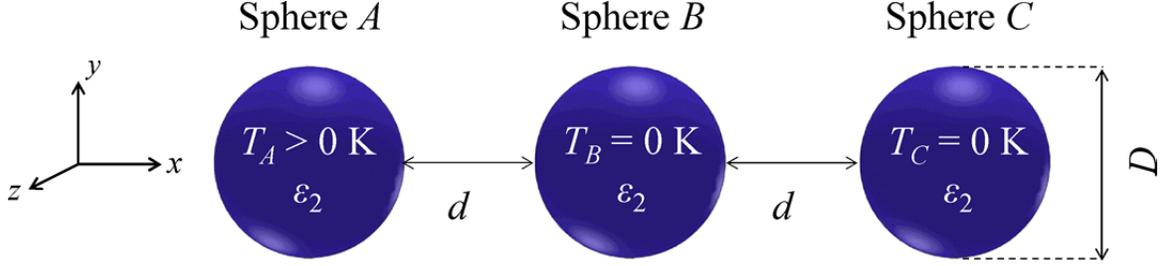

Figure 2. Schematic representation of the problem under consideration. All spheres have the same diameter $D$ and homogeneous dielectric function $\varepsilon_2$.

A closed-form expression for near-field thermal electromagnetic transport between three spheres is not available, except in the limit that each sphere can be modeled as an electric point dipole [35,36]. As such, the three-sphere problem is solved using the T-DDA in which objects are discretized into cubical sub-volumes conceptualized as electric point dipoles. Note that the convergence and accuracy of the T-DDA have been discussed in Ref. [44]. The total dipole moment $\mathbf{p}_i$ of sub-volume $i$ is the sum of an induced dipole moment $\mathbf{p}_i^{ind}$ and a fluctuating dipole moment $\mathbf{p}_i^{fl}$. By applying the discretization scheme described above, the volume integral equation for the electric field (Eq. (17)) can be written in terms of dipole moments [44]:

$$\frac{1}{\alpha_i}\mathbf{p}_i - \frac{k_0^2}{\varepsilon_0}\sum_{j\neq i}\overline{\overline{\mathbf{G}}}_{ij}\cdot\mathbf{p}_j = \frac{3}{(\varepsilon_{2,i}+2)}\frac{1}{\alpha_i^{CM}}\mathbf{p}_i^{fl} + \mathbf{E}_i^{inc} \qquad (20)$$

where $k_0$ is the magnitude of the wavevector in vacuum, while $\alpha_i$ and $\alpha_i^{CM}$ are respectively the radiative and Clausius-Mossotti polarizabilities. The ensemble average of the spatial correlation function of the fluctuating dipole moments is derived from the fluctuation-dissipation theorem given by Eq. (6):

$$\left\langle \mathbf{p}_i^{fl}(\omega) \otimes \mathbf{p}_i^{fl}(\omega') \right\rangle = \frac{4\varepsilon_0 \varepsilon''_{2,i} \Delta V_i}{\pi \omega} \Theta(\omega,T) \delta(\omega - \omega') \overline{\overline{\mathbf{I}}} \qquad (21)$$

where $\Delta V_i$ is the volume of sub-volume $i$. Equation (20) is a system of $3N$ scalar equations, where $N$ is the total number of sub-volumes in $V_2$, that can be written in a compact matrix form as follows:

$$\overline{\overline{\mathbf{A}}} \cdot \overline{\mathbf{P}} = \overline{\mathbf{E}}^{fdt} + \overline{\mathbf{E}}^{inc} \qquad (22)$$

where $\overline{\overline{\mathbf{A}}}$ is the interaction matrix, $\overline{\mathbf{E}}^{fdt}$ is a column vector containing the first term on the right-hand side of Eq. (20), $\overline{\mathbf{E}}^{inc}$ is the incident field column vector, while $\overline{\mathbf{P}}$ is the column vector of unknown total dipole moments [44]. The system of equations (22) is stochastic, and its direct solution provides the instantaneous total dipole moment in each sub-volume. The spectral power dissipated in the absorbers is calculated as:

$$\left\langle Q_{abs,\omega} \right\rangle = \frac{\omega}{2} \sum_{i \in abs} \left( \mathrm{Im}[(\alpha_i^{-1})^*] - \frac{2}{3} k_0^3 \right) \mathrm{tr} \left\langle \mathbf{p}_i^{ind} \otimes \mathbf{p}_i^{ind} \right\rangle \qquad (23)$$

where ergodicity is assumed [54,64]. In Eq. (23), $\mathrm{tr}\left\langle \mathbf{p}_i^{ind} \otimes \mathbf{p}_i^{ind} \right\rangle$ is the trace of the autocorrelation function of the induced dipole moment of sub-volume $i$. This term can be calculated directly from the system of equations (22) such that there is no need to compute the instantaneous total dipole moment, which is a quantity that is not experimentally observable [44].

It is assumed that sphere $A$ is emitting at a temperature $T_A$ of 300 K. The dimensionless sphere size $k_0 D$ and separation gap $d/\lambda$ are respectively 1.01 ($D$ = 1.6 μm) and 0.01 ($d$ = 100 nm), where the wavelength $\lambda$ is fixed at 10 μm. The study is performed for a resonant dielectric function $\varepsilon_2$ = -1.36 + 1.36$i$, corresponding to surface phonon-polariton resonance of a silica sphere, as well as for a non-resonant dielectric function $\varepsilon_2$ = 9 + 0.06$i$ representing a highly polarizable material with low losses. Note that the dipole approximation is not valid for this problem since the size of the spheres is much larger than the separation gap and of the same order of magnitude as the wavelength [44]. Figures 3 and 4 show cross sections of the spatial distribution of the volumetric power absorbed within sphere *B*, normalized by its maximum value, in the absence and in the presence of sphere *C* for the resonant and non-resonant dielectric functions, respectively. The cross sections are parallel to the *x-y* plane and pass through the center of the spheres. The number of sub-volumes used in the simulations is selected such that the error associated with the T-DDA in the absence of sphere *C* is no more than 2% when compared to the exact solution for two spheres [30-32]. The sub-volume size leading to a converged T-DDA solution depends mostly on the dielectric function of the spheres $\varepsilon_2$ and the sphere diameter to gap ratio *D*/*d* [44]. Since these parameters remain the same in the absence and in the presence of sphere *C*, the discretization used for the two-sphere geometry should lead to the same error for the three-sphere case. A fast convergence is observed for the resonant dielectric function, and an error of 2.0% is achieved in the absence of sphere *C* using 33552 uniform sub-volumes per sphere. For this discretization, the sub-volume size to gap ratio $\Delta$/*d* is 0.4. As explained in Ref. [44], the ratio $\Delta$/*d* leading to a converged T-DDA solution decreases as the dielectric function $\varepsilon_2$ increases. As such, a smaller $\Delta$/*d* value is required for the non-resonant dielectric function which is approximately 4.5 times larger than the resonant dielectric function. Non-uniform discretization is used for the non-resonant dielectric function in order to accelerate the convergence of the T-DDA. For this case, 27564 non-uniform sub-volumes per sphere result in an error of 1.0% in the absence of sphere *C*.

The ratio $\Delta/d$ for the non-uniform discretization increases from 0.08 at the front side of the spheres to 0.62 at the back side. A finer discretization is required at the front side of the spheres for an accurate representation of the variation of the gap along the $y$- and $z$-directions [44]. In the presence of sphere $C$, a fine discretization must also be applied to the back side of sphere $B$. In this case, 45800 non-uniform sub-volumes are used for discretizing sphere $B$.

For the resonant dielectric function, the power absorbed within sphere $B$ in the absence and presence of sphere $C$ is 198.6 nW/eV and 197.2 nW/eV, respectively. This slight decrease in the power absorbed is due to energy transfer from sphere $B$ to $C$ via tunneling of surface phonon-polaritons through the vacuum gap. Here, radiative energy transfer is dominated by surface phonon-polaritons [65,66], such that the contribution of propagating and frustrated modes to the power absorbed is negligible. The power absorbed within sphere $C$ is only 2.41 nW/eV, which is very small compared to the power absorbed by sphere $B$. The spatial distribution of the power absorbed within sphere $B$ is only slightly affected by the presence of sphere $C$. In the absence of sphere $C$, absorption within sphere $B$ decreases monotonically along the $x$-axis. In the presence of sphere $C$, a local minimum in the power absorbed is observed near the back side of sphere $B$. This local minimum is due to multiple scattering between spheres $B$ and $C$ leading to an increase of the power absorbed in the sub-volumes located immediately near the back side of sphere $B$.

For the non-resonant dielectric function, the power absorbed within sphere $B$ in the presence of sphere $C$ increases from $4.09\times10^{-4}$ nW/eV to $4.25\times10^{-4}$ nW/eV. The power absorbed within sphere $C$ is $2.85\times10^{-5}$ nW/eV. The slight increase in the power absorbed by sphere $B$ can be explained by the fact that energy transfer for the non-resonant dielectric function is solely due to propagating and frustrated modes. Since absorption is small for the non-resonant case, energy transfer from sphere $B$ to $C$ via tunneling of frustrated modes is modest. Propagating waves escaping from sphere *B* in the absence of sphere *C* are scattered back toward sphere *B* for the three-body problem. This phenomenon slightly increases the power absorbed within sphere *B* and induces a local maximum near the back side facing sphere *C*.

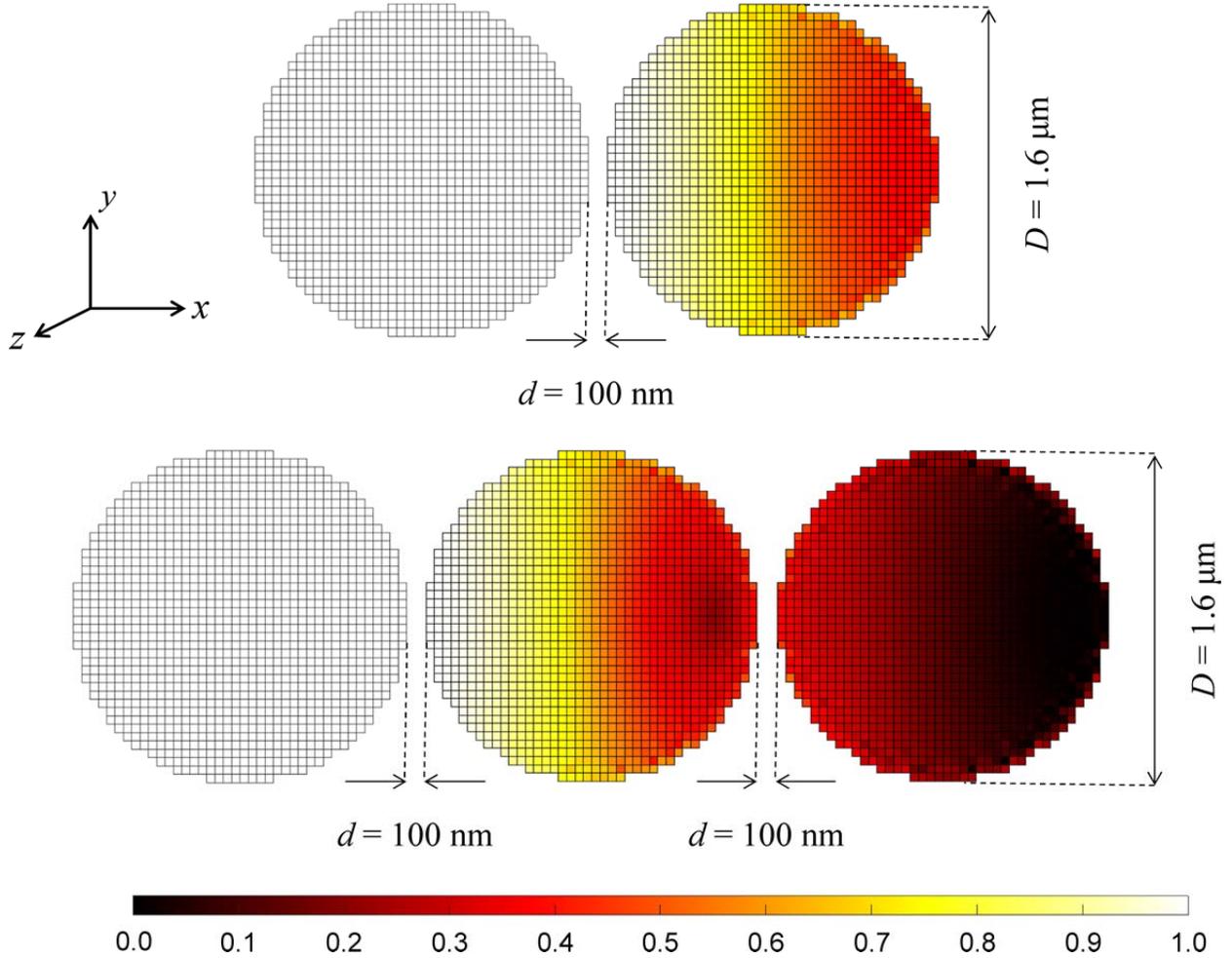

Figure 3. Spatial distribution of normalized volumetric power absorbed for the resonant dielectric function $\varepsilon_2 = -1.36 + 1.36i$ in the absence (top panel) and presence (bottom panel) of sphere *C*. For both the two- and three-sphere cases, 33552 uniform sub-volumes per sphere are used. The temperatures of spheres *A*, *B* and *C* are fixed at 300 K, 0 K and 0 K, respectively.

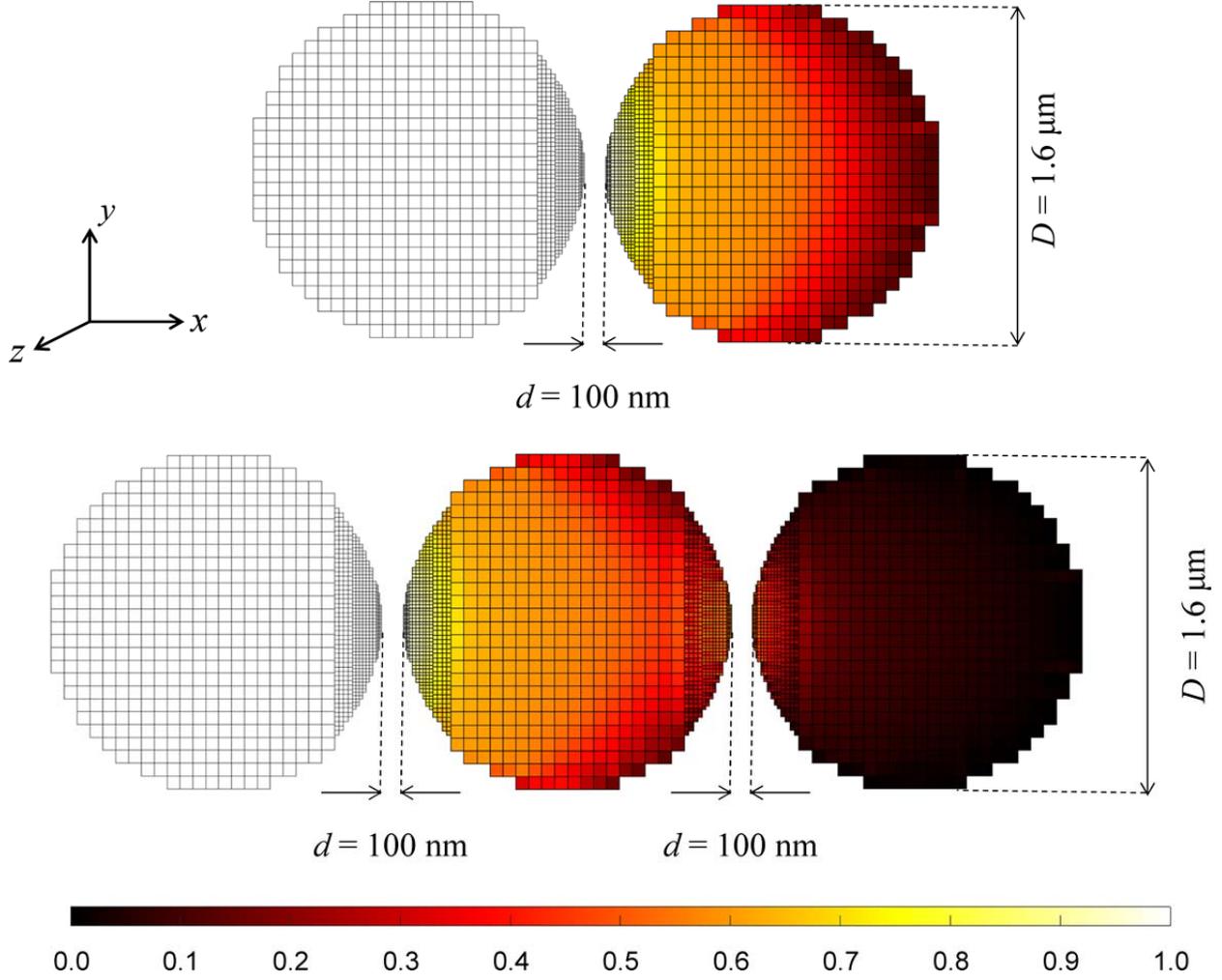

Figure 4. Spatial distribution of normalized volumetric power absorbed for the non-resonant dielectric function $\varepsilon_2 = 9 + 0.06i$ in the absence (top panel) and presence (bottom panel) of sphere $C$. For the two-sphere case, 27564 non-uniform sub-volumes per sphere are used. For the three-sphere case, spheres $A$ and $C$ are discretized into 27564 non-uniform sub-volumes while sphere $B$ is discretized into 45800 non-uniform sub-volumes. The temperatures of spheres $A$, $B$ and $C$ are fixed at 300 K, 0 K and 0 K, respectively.

It is also interesting to analyze a transient process in which the temperatures of spheres $A$ and $C$ are fixed at 300 K and 0 K, respectively. The temperature of sphere $B$, initially at 0 K, is calculated as a function of time using $\rho V c \cdot dT/dt = Q_{net,in}(t)$, where $Q_{net,in}$ is the net heat rate that account for thermal emission by sphere *B* at temperature $T_B$ calculated at time *t*. Here, it is assumed that sphere *B*, modeled with the thermophysical properties of silica ($k$ = 1.38 W/m·K, $\rho$ = 2220 kg/m$^3$, $c$ = 745 J/kg·K [67]), has a spatially uniform temperature at any time during the transient process even if the spatial distribution of volumetric power absorbed is non-uniform. This approximation is justified by the fact that the thermal resistance by conduction within sphere *B* is much smaller than the resistance by radiation. Figure 5 shows the temperature of sphere *B* as a function of time for both the resonant and non-resonant dielectric functions, and in the absence and presence of sphere *C*. For the calculations, the net heat rate $Q_{net,in}$ is integrated over a bandwidth of 0.02 eV.

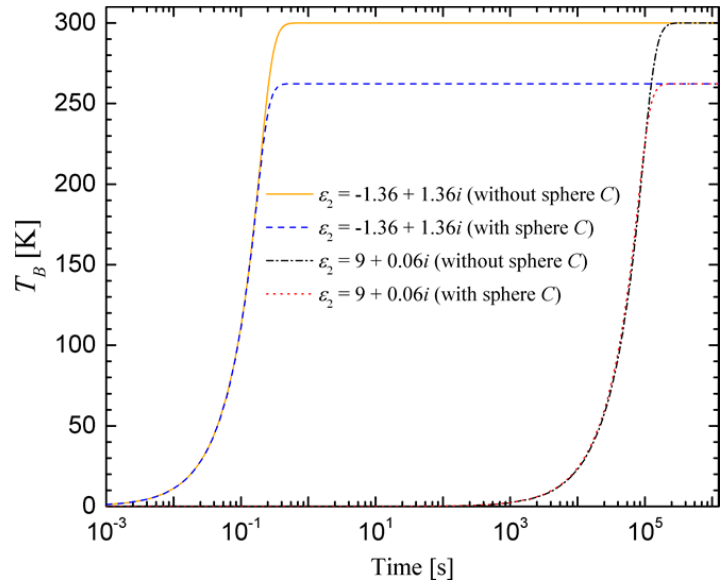

Figure 5. Temperature of sphere *B* as a function of time for the resonant ($\varepsilon_2$ = -1.36 + 1.36*i*) and non-resonant ($\varepsilon_2$ = 9 + 0.06*i*) dielectric functions, and in the absence and presence of sphere *C*. The temperatures of spheres *A* and *C* are fixed at 300 K and 0 K, respectively.

Regardless of the dielectric function, sphere *B* reaches a steady-state temperature of 262.3 K in the presence of sphere *C*. The time required to reach steady-state is approximately six orders of magnitude smaller at the resonant dielectric function, thus showing that effective localized radiative heating and cooling can be achieved via surface phonon-polaritons.

The analysis presented in this section could be extended to multiple spheres, which is important in the design of Mie resonance-based metamaterials for controlling thermal emission [61,68,69] and for understanding heating and cooling by near-field thermal radiation.

## 4. CONCLUSIONS

A near-field thermal electromagnetic transport formalism has been established by deriving a volume integral equation for the total electric field within the framework of fluctuational electrodynamics. The total electric field is the sum of an incident, a scattered and a fluctuating electric field, the latter being the thermal electromagnetic field generated by fluctuating currents. In the limiting case that thermal emission is negligible, it has been shown that the volume integral equation for the electric field reduces to the expression used in classical electromagnetic scattering. The near-field thermal electromagnetic transport formalism proposed in this paper can thus be seen as a generalization of the electromagnetic scattering theory.

Near-field radiative heat transfer is an emerging area of heat transfer physics and engineering. While a number of approaches have been proposed for modeling near-field thermal electromagnetic transport, the state-of-the-art is not as mature as in electromagnetic scattering. In this paper, an extension of the discrete dipole approximation, called the thermal discrete dipole approximation, has been used for analyzing a near-field thermal radiation problem involving three spheres starting from the volume integral equation for the total electric field. Other approaches used by the electromagnetic scattering community could be adapted to near-field thermal electromagnetic transport. In particular, an extension of the T-matrix method [70-72] to thermal radiation could provide a number of benchmark results for sophisticated problems involving, for instance, multiple spheres [73] and spheroids [74].

ACKNOWLEDGMENTS

This work was sponsored by the US Army Research Office under Grant no. W911NF-14-1-0210. The authors also acknowledge the Extreme Science and Engineering Discovery Environment (NSF grant no. ACI-1053575) and the Center for High Performance Computing at the University of Utah for providing the computational resources used in this study.

[22] B.J. Lee, C.J. Fu, and Z.M. Zhang, Appl. Phys. Lett. **87**, 071904 (2005).

[23] E. Rephaeli, A. Raman, and S. Fan, Nano Lett. **13**, 1457 (2013).

[24] C.R. Otey, W.T. Lau, and S. Fan, Phys. Rev. Lett. **104**, 154301 (2010).

[25] S. Basu and M. Francoeur, Appl. Phys. Lett. **98**, 113106 (2011).

[26] P. Ben-Abdallah and S.-A. Biehs, Phys. Rev. Lett. 112, 044301 (2014).

[27] P. Ben-Abdallah and S.-A. Biehs, AIP Adv. **5**, 053502 (2015).

[28] A. Narayanaswamy and G. Chen, Annu. Rev. Heat Transfer **14**, 169 (2005).

[29] M. Francoeur, M.P. Mengüç, and R. Vaillon, J. Quant. Spectrosc. Radiat. Transfer **110**, 2002 (2009).

[30] A. Narayanaswamy and G. Chen, Phys. Rev. B **77**, 075125 (2008).

[31] K. Sasihithlu and A. Narayanaswamy, Opt. Express **19**, 772 (2011).

[32] K. Sasihithlu and A. Narayanaswamy, Phys. Rev. B **83**, 161406(R) (2011).

[33] M. Krüger, T. Emig, and M. Kardar, Phys. Rev. Lett. **106**, 210404 (2011).

[34] C. Otey and S. Fan, Phys. Rev. B **84**, 245431 (2011).

[35] P. Ben-Abdallah, S.-A. Biehs, and K. Joulain, Phys. Rev. Lett. **107**, 114301 (2011).

[36] R. Messina, M. Tschikin, S.-A. Biehs, and P. Ben-Abdallah, Phys. Rev. B **88**, 104307 (2013).
21